\documentclass[proceedings]{JHEP3}
\usepackage{eurosym}
\usepackage{amssymb}
\usepackage{amsfonts}
\usepackage{amsmath,epsfig}
\usepackage{graphicx}

\newbox\mybox

\newcommand\fverb{\setbox\mybox=\hbox\bgroup\verb}
\newcommand\fverbdo{\egroup\medskip\noindent\fbox{\unhbox\mybox}\ }
\newcommand\fverbit{\egroup\item[\fbox{\unhbox\mybox}]}
\conference{QES quantum systems with explicitly time-dependent Hamiltonians}
\abstract{For a large class of time-dependent non-Hermitain Hamiltonians expressed in terms linear and bilinear 
combinations of the generators for an Euclidean Lie-algebra respecting different types of PT-symmetries, 
we find explicit solutions to the time-dependent Dyson equation. A specific Hermitian model with explicit time-dependence is analyzed further and shown to 
be quasi-exactly solvable. Technically we constructed the Lewis-Riesenfeld invariants making use
of the metric picture, which is an equivalent alternative to the
Schr\"{o}dinger, Heisenberg and interaction picture containing the time-dependence in the
metric operator that relates the time-dependent Hermitian Hamiltonian to a static
non-Hermitian Hamiltonian.}

\title{Quasi-exactly solvable quantum systems with explicitly time-dependent
Hamiltonians}
\author{Andreas Fring and Thomas Frith \\
Department of Mathematics, City University London,\\
Northampton Square, London EC1V 0HB, UK \\
E-mail: a.fring@city.ac.uk, thomas.frith@city.ac.uk}

\input{tcilatex}
\begin{document}

\section{Introduction}

Quasi-exactly solvable (QES) quantum systems are characterized by the
feature that only part of their infinite energy spectrum and corresponding
eigenfunctions can be calculated analytically. Systematic studies of such
type of systems have been carried out by casting them into the form of Lie
algebraic quantities \cite{OP2,OP4} and making use of the property that the
eigenfunctions of the corresponding Hamiltonian systems form a flag which
coincides with the finite dimensional representation space of the associated
Lie algebras. QES systems that can be cast into such a form are usually
referred to as QES models of Lie algebraic type \cite{Turbiner00,Tur0}. The
relevant underlying algebras are either of $sl_{2}(\mathbb{C})$-type, with
their compact and non-compact real forms $su(2)$ and $su(1,1)$, respectively 
\cite{Hum}, or of Euclidean Lie algebras type \cite%
{E2Fring,E2Fring2,fring2016unifying}. The latter class was found to be
particularly useful when dealing with certain types of non-Hermitian systems.

While many QES models have been studied in stationary settings, little is
known for time-dependent systems. So far a time-dependence has only been
introduced into the eigenfunctions in form of a dynamical phase \cite%
{mayer2000time,hou1999quasi}. However, no QES systems with explicitly
time-dependent Hamiltonians have been considered up to now. The main purpose
of this article is to demonstrate how they can be dealt with and to initiate
further studies of such type of systems. We provide the analytical solutions
to a QES Hamiltonian quantum system with explicit time-dependence. As a
concrete example we consider QES systems of $E_{2}$-Lie algebraic type.
Technically we make use of the metric picture \cite{AndTom1,AndTom2}, which
is an alternative to the Schr\"{o}dinger, Heisenberg and interaction
picture. It will allow us to solve a Hermitian time-dependent Hamiltonian
system by solving first a static non-Hermitian system as an auxiliary
problem with a time-dependence in the metric operator.

The Hermitian Hamiltonian systems we study here are of the general form%
\begin{equation}
h(t)=\mu _{JJ}(t)J^{2}+\mu _{J}(t)J+\mu _{u}(t)u+\mu _{v}(t)v+\mu
_{uu}(t)u^{2}+\mu _{vv}(t)v^{2}+\mu _{uv}(t)uv,  \label{1}
\end{equation}%
where the time-dependent coefficient functions $\mu _{i}$, $i\in
\{J,JJ,u,v,uu,vv,uv\}$, are real and $u$, $v$ and $J$ denote the three
generators that span the Euclidean-algebra $E_{2}$. They obey the
commutation relations%
\begin{equation}
\left[ u,J\right] =iv,\qquad \left[ v,J\right] =-iu,\qquad \text{and\qquad }%
\left[ u,v\right] =0.  \label{E2}
\end{equation}%
Considering here only Hermitian representations with $J^{\dagger }=J$, $%
v^{\dagger }=v$ and $u^{\dagger }=u$, the Hamiltonian in equation (\ref{1})
is clearly Hermitian. Standard representation are for instance the
trigonometric representation $J:=-i\partial _{\theta }$, $u:=\sin \theta $
and $v:=\cos \theta $ or a two-dimensional representation $J:=yp_{x}-xp_{y}$%
, $u:=x$ or $v:=y$ with $x$, $y$, $p_{x}$, $p_{y}$ denoting Heisenberg
canonical variables with non-vanishing commutators $\left[ x,p_{x}\right] =%
\left[ y,p_{y}\right] =i$. We have set here and mostly in what follows to $%
\hbar =1$.

We briefly recall from \cite{AndTom1,AndTom2} what is meant by the \textit{%
metric picture}. It is well known that the Schr\"{o}dinger and the
Heisenberg picture are equivalent with the former containing the
time-dependence entirely in the states and the latter entirely in the
operators. $\mathcal{PT}$-symmetric/quasi-Hermitian systems \cite%
{Urubu,Benderrev,Alirev} allow for yet another equivalent variant in which
the time-dependence is contained entirely in the metric operator.\ In order
to see that we first need to solve the time-dependent Dyson relation \cite%
{CA,time1,time6,time7,fringmoussa,AndTom1,AndTom2,AndTom3,AndTom4,mostafazadeh2018energy}
which in general reads 
\begin{equation}
h(t)=\eta (t)H(t)\eta ^{-1}(t)+i\hbar \partial _{t}\eta (t)\eta ^{-1}(t),
\label{TDDE}
\end{equation}%
involving a time-dependent non-Hermitian Hamiltonian $H(t)\neq H^{\dagger
}(t)$ and the Dyson operator $\eta $ related to the metric operator $\rho $
as $\rho =\eta ^{\dagger }\eta $. For our purposes we will eventually take
the Hamiltonian to be time-independent $H(t)\rightarrow H$, with $h(t)$
satisfying the time-dependent Schr\"{o}dinger equation $h(t)\phi (t)=i\hbar
\partial _{t}\phi (t)$ and $H$ the time-independent Schr\"{o}dinger equation 
$H\psi =E\psi $ with energy eigenvalue $E$. The corresponding wavefunctions
are related as $\phi (t)=\eta (t)\psi $.

Before we solve a concrete system in a quasi-exactly solvable fashion we
consider first the fully time-dependent Dyson relation with time-dependent
non-Hermitian Hamiltonian $H(t)$ and investigate which type of Hamiltonians
can be related to the Hermitian Hamiltonian $h(t)$ in (\ref{1}). We will see
that in some cases we are even forced to take $H(t)$ or part of it to be
time-independent. As not many explicit solutions to the time-dependent Dyson
relation are known, this will be a valuable result in itself. 

Our manuscript is organized as follows: In section 2 we explore various
types of $\mathcal{PT}$-symmetries that leave the Euclidean $E_{2}$-algebra
invariant and investigate time-dependent non-Hermitian Hamiltonians in terms 
$E_{2}$-algebraic generators that respect these symmetries. We find new
solutions to the time-dependent Dyson relation for those type of
Hamiltonians by computing the corresponding Hermitian Hamiltonians and the
Dyson map. In section 3 we provide analytical solutions for a concrete model
respecting a particular $\mathcal{PT}$-symmetry. We compute the eigenstates
of the Lewis-Riesenfeld invariants and the time-dependent Hermitian
Hamiltonian in a quasi-exactly solvable fashion. A three-level system is
presented in more detail. Our conclusions are stated in section 4.

\section{Solutions to the time-dependent Dyson equation for $E_{2}$%
-Hamiltonians}

A key property in the study and classification of Hamiltonian systems
related to the $E_{2}$-algebra are the antilinear symmetries \cite{EW} that
leave the algebra (\ref{E2}) invariant. Given the general context of $%
\mathcal{PT}$-symmetric/quasi-Hermitian systems we call these symmetries $%
\mathcal{PT}_{i},i=1,2,\ldots $ As discussed in more detail in \cite%
{DFM,DFM2}, there are many options which all give rise to models with
qualitatively quite distinct features. It is easy to see that each of the
following antilinear maps leave all the commutation relations (\ref{E2})
invariant 
\begin{equation}
\begin{array}{lllll}
\mathcal{PT}_{1}:~~ & J\rightarrow -J,~~ & u\rightarrow -u,~~ & v\rightarrow
-v,~~~ & i\rightarrow -i, \\ 
\mathcal{PT}_{2}: & J\rightarrow -J, & u\rightarrow u, & v\rightarrow v, & 
i\rightarrow -i, \\ 
\mathcal{PT}_{3}: & J\rightarrow J, & u\rightarrow v, & v\rightarrow u, & 
i\rightarrow -i, \\ 
\mathcal{PT}_{4}: & J\rightarrow J, & u\rightarrow -u, & v\rightarrow v, & 
i\rightarrow -i, \\ 
\mathcal{PT}_{5}: & J\rightarrow J, & u\rightarrow u, & v\rightarrow -v, & 
i\rightarrow -i.%
\end{array}
\label{PT}
\end{equation}%
Next we seek non-Hermitian Hamiltonians that respect either of these
symmetries. Focussing here on time-dependent Hamiltonians consisting
entirely of linear and bilinear combinations of $E_{2}$-generators they can
all be cast into the general form 
\begin{eqnarray}
H_{\mathcal{PT}_{i}}(t) &=&\mu _{JJ}(t)J^{2}+\mu _{J}(t)J+\mu _{u}(t)u+\mu
_{v}(t)v+\mu _{uJ}(t)uJ+\mu _{vJ}(t)vJ  \label{H} \\
&&+\mu _{uu}(t)u^{2}+\mu _{vv}(t)v^{2}+\mu _{uv}(t)uv.  \notag
\end{eqnarray}%
Demanding that $\left[ H_{\mathcal{PT}_{i}}(t),\mathcal{PT}_{i}\right] =0$,
the symmetries are implemented by taking the coefficient functions to be
either real, purely imaginary or relate different functions to each other by
conjugation. For the different symmetries in (\ref{PT}) we are forced to
take 
\begin{equation}
\begin{array}{lll}
\mathcal{PT}_{1}:~~ & (\mu _{J},\mu _{u},\mu _{v})\in i\mathbb{R},~~ & (\mu
_{JJ},\mu _{uJ},\mu _{vJ},\mu _{uu},\mu _{vv},\mu _{uv})\in \mathbb{R},~~ \\ 
\mathcal{PT}_{2}: & (\mu _{J},\mu _{uJ},\mu _{vJ})\in i\mathbb{R}, & (\mu
_{u},\mu _{v},\mu _{JJ},\mu _{uu},\mu _{vv},\mu _{uv})\in \mathbb{R}, \\ 
\mathcal{PT}_{3}: & (\mu _{JJ},\mu _{J},\mu _{uv})\in \mathbb{R}, & \mu
_{u}=\mu _{v}^{\ast },\mu _{uJ}=\mu _{vJ}^{\ast },\mu _{uu}=\mu _{vv}^{\ast }
\\ 
\mathcal{PT}_{4}: & (\mu _{u},\mu _{uJ},\mu _{uv})\in i\mathbb{R}, & (\mu
_{J},\mu _{v},\mu _{JJ},\mu _{vJ},\mu _{uu},\mu _{vv})\in \mathbb{R}, \\ 
\mathcal{PT}_{5}: & (\mu _{v},\mu _{vJ},\mu _{uv})\in i\mathbb{R},~ & (\mu
_{J},\mu _{u},\mu _{JJ},\mu _{uJ},\mu _{uu},\mu _{vv})\in \mathbb{R}.%
\end{array}
\label{coe}
\end{equation}%
Except for very specific combinations of the coefficient functions, the
Hamiltonians $H_{\mathcal{PT}_{i}}(t)$ are non-Hermitian in general.

We now solve the time-dependent Dyson relation (\ref{TDDE}) for $\eta (t)$
by mapping different $\mathcal{PT}_{i}$-symmetric versions of $H(t)$ to a
Hermitian Hamiltonian $h(t)$ of the form (\ref{1}). For the time-dependent
Dyson map we make an Ansatz in terms of all the $E_{2}$-generators 
\begin{equation}
\eta (t)=e^{\tau (t)v}e^{\lambda (t)J}e^{\rho (t)u}.  \label{eta}
\end{equation}%
At this point we allow $\lambda ,\tau ,\rho \in \mathbb{C}$, keeping in mind
that $\eta (t)$ does not have to be Hermitian. We exclude here unitary
operators, i.e. $\lambda ,\tau ,\rho \in i\mathbb{R}$, as in that case $\eta
(t)$ just becomes a gauge transformation. The adjoint action of this
operator on the $E_{2}$-generators is computed by using the standard
Baker-Campbell-Haussdorff formula 
\begin{eqnarray}
\eta J\eta ^{-1} &=&J+i\rho \cosh (\lambda )v-[i\tau +\rho \sinh (\lambda
)]u,  \label{ad1} \\
\eta u\eta ^{-1} &=&\cosh (\lambda )u-i\sinh (\lambda )v,  \label{ad2} \\
\eta v\eta ^{-1} &=&\cosh (\lambda )v+i\sinh (\lambda )u.  \label{ad3}
\end{eqnarray}%
The gauge-like term in (\ref{TDDE}) acquires the form%
\begin{equation}
i\dot{\eta}\eta ^{-1}=i\dot{\lambda}J+\left[ i\dot{\rho}\cosh \left( \lambda
\right) +\tau \dot{\lambda}\right] u+\left[ \dot{\rho}\sinh \left( \lambda
\right) +i\dot{\tau}\right] v.  \label{gt}
\end{equation}%
As common, we abbreviate here time-derivatives by overdots. For the
computation of the time-dependent energy operator $\tilde{H}(t)$, see below,
we also require the term%
\begin{equation}
i\eta ^{-1}\dot{\eta}=i\dot{\lambda}J+\left[ i\dot{\rho}+\dot{\tau}\sinh
\left( \lambda \right) \right] u+\left[ \rho \dot{\lambda}+i\dot{\tau}\cosh
\left( \lambda \right) \right] v.  \label{auxen}
\end{equation}%
Using (\ref{ad1})-(\ref{ad3}) we calculate next the adjoint action of $\eta $
on $H(t)$ and add the expression in (\ref{gt}). Demanding that the result is
Hermitian will constrain the time-dependent functions $\mu _{i}(t)$, $%
\lambda (t)$, $\tau (t)$ and $\rho (t)$. We need to treat each $\mathcal{PT}$%
-symmetry separately.

\subsection{Time-dependent $\mathcal{PT}_{1}$-invariant Hamiltonians}

For convenience we take the coefficient function $\mu _{JJ}$ to be
time-independent. For the $\mathcal{PT}_{1}$-invariant Hamiltonian with
coefficient functions as specified in (\ref{coe}) we have to be aware that
for $\mu _{J}=\mu _{uJ}=$ $\mu _{vJ}=0$ the Hamiltonian $H_{\mathcal{PT}%
_{1}}(t)$ becomes Hermitian. Substituting the general form for $H_{\mathcal{%
PT}_{1}}(t)$ into (\ref{TDDE}), using (\ref{ad1})-(\ref{ad3}), (\ref{gt}),
reading off the coefficients in front of the generators and demanding that
the right hand side becomes Hermitian enforces to take the functions $%
\lambda ,\tau ,\rho \in \mathbb{R}$ in (\ref{eta}). The resulting Hermitian
Hamiltonian is%
\begin{eqnarray}
h_{\mathcal{PT}_{1}} &=&J^{2}\mu _{JJ}+\frac{\left[ \mu _{vJ}\tanh \lambda
-\mu _{J}\mu _{vJ}\right] \sinh \lambda }{2\mu _{JJ}}u-\frac{\mu _{J}\mu
_{uJ}\tanh \lambda \func{sech}\lambda }{2\mu _{JJ}}v \\
&&+\left( \mu _{uu}-\frac{\mu _{uJ}^{2}\tanh ^{2}\lambda }{4\mu _{JJ}}%
\right) u^{2}+\left( \mu _{uu}+\frac{\cosh ^{2}(\lambda )\mu _{vJ}^{2}-\mu
_{uJ}^{2}}{4\mu _{JJ}}\right) v^{2}+\mu _{uv}uv,  \notag \\
&&+\frac{\mu _{uJ}}{2}\func{sech}\lambda \{u,J\}+\frac{\mu _{vJ}}{2}\cosh
\lambda \{v,J\}  \notag
\end{eqnarray}%
with 7 constraining relations%
\begin{eqnarray}
\lambda &=&-\dint\nolimits^{t}\mu _{J}(s)ds,~~\tau =\frac{\mu _{vJ}\sinh
\lambda }{2\mu _{JJ}},~~\rho =\frac{\mu _{uJ}\tanh \lambda }{2\mu _{JJ}}%
,~~\mu _{vv}=\mu _{uu}+\frac{\mu _{vJ}^{2}-\mu _{uJ}^{2}}{4\mu _{JJ}}, \\
\mu _{uv} &=&\frac{\mu _{uJ}\mu _{vJ}}{2\mu _{JJ}},~~\mu _{u}=\frac{\mu
_{J}\mu _{uJ}-\dot{\mu}_{uJ}\tanh \lambda }{2\mu _{JJ}}+\frac{\mu _{vJ}}{2}%
,~~\mu _{v}=\frac{\mu _{J}\mu _{vJ}-\dot{\mu}_{vJ}\tanh \lambda }{2\mu _{JJ}}%
-\frac{\mu _{uJ}}{2}.  \notag
\end{eqnarray}%
Thus from the original 12 free parameters, i.e. the 9 coefficient functions $%
\mu _{i}$ and the 3 functions $\lambda ,\tau ,\rho $ in the Dyson map, we
can still freely choose 5. In comparison with the other $\mathcal{PT}_{i}$%
-symmetries, this is the most constrained case. We also note that this
system is the only one in which all three functions in the Dyson map are
constrained when we take the coefficient functions $\mu _{i}$ as primary
quantities.

\subsection{Time-dependent $\mathcal{PT}_{2}$-invariant Hamiltonians}

The Hamiltonian $H_{\mathcal{PT}_{2}}(t)$ becomes Hermitian for $\mu _{J}=0$%
, $\mu _{uJ}=2\mu _{u}$, $\mu _{vJ}=-2\mu _{u}$, but is non-Hermitian
otherwise. Preceding as in the previous section the implementation of (\ref%
{TDDE}) enforces to take $\tau ,\rho \in \mathbb{R}$ and $\lambda \in i%
\mathbb{R}$ in (\ref{eta}), which makes the Dyson map $\mathcal{PT}_{2}$%
-symmetric. The Hermitian Hamiltonian is computed to 
\begin{eqnarray}
h_{\mathcal{PT}_{2}} &=&\mu _{JJ}J^{2}+\dot{\lambda}J+\left[ \left( \mu _{u}+%
\frac{\mu _{vJ}}{2}\right) \cos \lambda +\left( \frac{\mu _{uJ}}{2}-\mu
_{v}\right) \sin \lambda \right] u \\
&&+\left[ \left( \mu _{v}-\frac{\mu _{uJ}}{2}\right) \cos \lambda +\left(
\mu _{u}+\frac{\mu _{vJ}}{2}\right) \sin \lambda \right] v+\left[ \left( 
\frac{\mu _{uJ}^{2}-\mu _{vJ}^{2}}{8\mu _{JJ}}+\frac{\mu _{uu}-\mu _{vv}}{2}%
\right) \cos (2\lambda )\right.  \notag \\
&&-\left. \left( \frac{\mu _{uJ}\mu _{vJ}}{4\mu _{JJ}}+\frac{\mu _{uv}}{2}%
\right) \sin (2\lambda )+\frac{\mu _{uJ}^{2}+\mu _{vJ}^{2}}{8\mu _{JJ}}+%
\frac{\mu _{uu}+\mu _{vv}}{2}\right] u^{2}  \notag \\
&&+\left[ \left( \frac{\mu _{uJ}^{2}}{4\mu _{JJ}}+\mu _{uu}\right) \sin
^{2}\lambda +\left( \frac{\mu _{uJ}\mu _{vJ}}{4\mu _{JJ}}+\frac{\mu _{uv}}{2}%
\right) \sin 2\lambda +\left( \frac{\mu _{vJ}^{2}}{4\mu _{JJ}}+\mu
_{vv}\right) \cos ^{2}\lambda \right] v^{2}  \notag \\
&&+\left[ \left( \frac{\mu _{uJ}^{2}-\mu _{vJ}^{2}}{4\mu _{JJ}}+\mu
_{uu}-\mu _{vv}\right) \sin (2\lambda )+\left( \frac{\mu _{uJ}\mu _{vJ}}{%
2\mu _{JJ}}+\mu _{uv}\right) \cos (2\lambda )\right] uv,  \notag
\end{eqnarray}%
with 5 constraining relations 
\begin{equation}
\tau =\frac{\mu _{uJ}}{2\mu _{JJ}}\sec \lambda ,\qquad \rho =-\frac{\mu
_{vJ}+\mu _{uJ}\tan \lambda }{2\mu _{JJ}},\qquad \mu _{J}=\dot{\mu}_{uJ}=%
\dot{\mu}_{vJ}=0.  \label{2con}
\end{equation}%
We note that we have less constraints as in the previous section, but some
of the coefficient functions can no longer be taken to be time-dependent and
one even has to vanish. One of the three functions in the Dyson map, e.g. $%
\lambda $, can be freely chosen. Compared to the other cases this is the
only one for which $\eta $ has the same $\mathcal{PT}_{i}$-symmetry as the
corresponding non-Hermitian Hamiltonian $H_{\mathcal{PT}_{i}}(t)$ when
taking the constraints on $\tau ,\rho ,\lambda $ into account.

\subsection{Time-dependent $\mathcal{PT}_{3}$-invariant Hamiltonians}

The Hamiltonian $H_{\mathcal{PT}_{3}}(t)$ becomes Hermitian for $\mu
_{vJ}=\mu _{uu}=0$ and $\mu _{uJ}=2\mu _{v}$. Using the same arguments as
above, we are forced to take $\tau ,\rho \in \mathbb{R}$ and $\lambda \in i%
\mathbb{R}$ in (\ref{eta}). The Hermitian Hamiltonian is computed to 
\begin{eqnarray}
h_{\mathcal{PT}_{3}} &=&J^{2}\mu _{JJ}+\left( \mu _{J}-\text{$\dot{\lambda}$}%
\right) J+\cos \lambda \left( \mu _{u}-\frac{\mu _{vJ}}{2}\right) (u+v)+\sin
\lambda \left( \mu _{u}-\frac{\mu _{vJ}}{2}\right) (v-u)~~~~~~~~ \\
&&+\left( \mu _{vv}+\frac{\mu _{vJ}^{2}}{4\mu _{JJ}}\right) \left(
u^{2}+v^{2}\right) +\left( \frac{\mu _{vJ}^{2}}{4\mu _{JJ}}-\frac{\mu _{uv}}{%
2}\right) \sin (2\lambda )\left( u^{2}-v^{2}\right)  \notag \\
&&+\frac{\mu _{uJ}}{2}\cos \lambda \left[ \{v,J\}+\{u,J\}\right] +\frac{\mu
_{uJ}}{2}\sin \lambda \left[ \{v,J\}-\{u,J\}\right]  \notag \\
&&+\cos (2\lambda )\left( \mu _{uv}-\frac{\mu _{vJ}^{2}}{2\mu _{JJ}}\right)
uv,  \notag
\end{eqnarray}%
with 5 constraining relations 
\begin{equation}
\tau =\frac{\mu _{vJ}}{2\mu _{JJ}}\sec \lambda ,~~\rho =\frac{\mu _{vJ}-\mu
_{vJ}\tan \lambda }{2\mu _{JJ}},~~\mu _{v}=\frac{\mu _{vJ}}{2}+\frac{\mu
_{J}\mu _{vJ}}{2\mu _{JJ}},~~\mu _{uv}=-\frac{\mu _{vJ}\mu _{uJ}}{2\mu _{JJ}}%
,~~\dot{\mu}_{vJ}=0.
\end{equation}%
Once again one of the coefficient functions has to be time-independent and
one of the three functions in the Dyson map can be chosen freely.

\subsection{Time-dependent $\mathcal{PT}_{4}$-invariant Hamiltonians}

The Hamiltonian $H_{\mathcal{PT}_{4}}(t)$ becomes Hermitian for $\mu
_{uJ}=\mu _{uv}=0$ and $\mu _{vJ}=2\mu _{u}$. By the same reasoning as above
we have to take $\tau ,\rho \in \mathbb{R}$ and $\lambda \in i\mathbb{R}$ in
(\ref{eta}). The Hermitian Hamiltonian results to to 
\begin{eqnarray}
h_{\mathcal{PT}_{4}} &=&J^{2}\mu _{JJ}+\left( \mu _{J}-\text{$\dot{\lambda}$}%
\right) J+\sin \lambda \left( \frac{\mu _{uJ}}{2}-\mu _{v}\right) u+\cos
\lambda \left( \mu _{v}-\frac{\mu _{uJ}}{2}\right) v \\
&&+\left( \mu _{uu}-\mu _{vv}+\frac{\mu _{uJ}^{2}}{4\mu _{JJ}}\right) \sin
(2\lambda )uv-\frac{\mu _{vJ}}{2}\sin \lambda \{u,J\}+\frac{\mu _{vJ}}{2}%
\cos \lambda \{v,J\}  \notag \\
&&+\left[ \left( \frac{\mu _{uu}-\mu _{vv}}{2}+\frac{\mu _{uJ}^{2}}{8\mu
_{JJ}}\right) \cos (2\lambda )+\left( \frac{\mu _{uu}+\mu _{vv}}{2}\right) +%
\frac{\mu _{uJ}^{2}}{8\mu _{JJ}}\right] u^{2}  \notag \\
&&+\left[ \left( \mu _{uu}+\frac{\mu _{uJ}^{2}}{4\mu _{JJ}}\right) \sin
^{2}\lambda +\cos ^{2}\lambda \mu _{vv}\right] v^{2},  \notag
\end{eqnarray}%
with 5 constraining relations 
\begin{equation}
\tau =\frac{\mu _{uJ}}{2\mu _{JJ}}\sec \lambda ,\quad \rho =-\frac{\mu
_{uJ}\tan \lambda }{2\mu _{JJ}},\quad \mu _{u}=\frac{\mu _{vJ}}{2}+\frac{\mu
_{J}\mu _{uJ}}{2\mu _{JJ}},\quad \mu _{uv}=\frac{\mu _{vJ}\mu _{uJ}}{2\mu
_{JJ}},\quad \dot{\mu}_{uJ}=0.
\end{equation}%
This case is similar to the previous one with one of the coefficient
functions forced to be time-independent and one of the three functions in
the Dyson map being freely choosable.

\subsection{Time-dependent $\mathcal{PT}_{5}$-invariant Hamiltonians}

The Hamiltonian $H$ becomes Hermitian for $\mu _{vJ}=\mu _{uv}=0$ and $\mu
_{uJ}=-2\mu _{v}$. Here we have to take $\rho \in \mathbb{R}$ and $\lambda
,\tau \in i\mathbb{R}$ in (\ref{eta}). The Hermitian Hamiltonian is computed
to 
\begin{eqnarray}
h_{\mathcal{PT}_{5}} &=&J^{2}\mu _{JJ}+\left( \mu _{J}-\text{$\dot{\lambda}$}%
\right) J+\left( \tau \mu _{J}+\frac{\mu _{uJ}}{2}\cos \lambda \right)
\{u,J\}+\frac{\mu _{uJ}}{2}\sin \lambda \{v,J\} \\
&&+\left[ \tau \left( \mu _{J}-\text{$\dot{\lambda}$}\right) +\cos \lambda
\left( \mu _{u}+\frac{\mu _{vJ}}{2}\right) \right] u+\left[ \sin \lambda
\left( \mu _{u}+\frac{\mu _{vJ}}{2}\right) -\dot{\tau}\right] v  \notag \\
&&+\left[ \tau ^{2}\mu _{JJ}+\sin ^{2}\lambda \left( \frac{\mu _{vJ}^{2}}{%
4\mu _{JJ}}+\mu _{vv}\right) +\tau \cos \lambda \mu _{uJ}+\cos ^{2}\lambda
\mu _{uu}\right] u^{2}  \notag \\
&&+\sin \lambda \left[ 2\cos \lambda \left( \mu _{uu}-\mu _{vv}-\frac{\mu
_{vJ}^{2}}{4\mu _{JJ}}\right) +\tau \tau \mu _{uJ}\right] uv  \notag \\
&&+\left[ \left( \frac{\mu _{vJ}^{2}}{4\mu _{JJ}}+\mu _{vv}\right) \cos
^{2}\lambda +\mu _{uu}\sin ^{2}\lambda \right] v^{2},  \notag
\end{eqnarray}%
with only 4 constraining relations 
\begin{equation}
\rho =-\frac{\mu _{vJ}}{2\mu _{JJ}},\quad \mu _{v}=-\frac{\mu _{uJ}}{2}+%
\frac{\mu _{J}\mu _{vJ}}{2\mu _{JJ}},~~\dot{\mu}_{vJ}=0,~~\mu _{uv}=\frac{%
\mu _{vJ}\mu _{uJ}}{2\mu _{JJ}}.
\end{equation}%
In comparison with the other symmetries, this is the least constraint case.
From the three functions in the Dyson map only one is constraint and the
others can be chosen freely. However, one of the coefficient functions needs
to be time-independent.

\section{Time-dependent quasi-exactly solvable systems}

We will now specify one particular model and show how it can be
quasi-exactly solved in the metric picture. Since the $\mathcal{PT}_{2}$
symmetry appears to be somewhat special, in the sense that it is the only
case for which the Dyson map respects the same symmetry as the Hamiltonian,
we consider a particular non-Hermitian $\mathcal{PT}_{2}$-symmetric
time-independent Hamiltonian of the form

\begin{equation}
\hat{H}=m_{JJ}J^{2}+m_{v}v+m_{vv}v^{2}+im_{uJ}uJ.  \label{HH}
\end{equation}%
Given the constraining equations (\ref{2con}), we could in principle take $%
m_{v}$, $m_{vv}$ to be time dependent, but to enforce the metric picture we
take here all four coefficients $m_{JJ}$, $m_{v}$, $m_{vv}$ and $m_{uJ}$ to
be time-independent real constants. According to the analysis in section
2.2, the time-dependent Dyson map%
\begin{equation}
\eta (t)=e^{\tau (t)v}e^{i\lambda (t)J}e^{\varrho (t)u},\quad \tau (t)=\frac{%
\mu _{uJ}}{2\mu _{JJ}}\sec \lambda (t),\quad \varrho (t)=-\frac{\mu _{uJ}}{%
2\mu _{JJ}}\tan \lambda (t),  \label{etas}
\end{equation}%
with $\lambda ,\tau ,\rho \in \mathbb{R}$, maps the time-independent
non-Hermitian Hamiltonian $\hat{H}$ to the time-dependent Hermitian
Hamiltonian%
\begin{eqnarray}
\hat{h}(t) &=&m_{JJ}J^{2}-\dot{\lambda}J+\sin \lambda \left( \frac{m_{uJ}}{2}%
-m_{v}\right) u+\cos \lambda \left( m_{v}-\frac{m_{uJ}}{2}\right) v
\label{hh} \\
&&+\left[ \cos (2\lambda )\left( \frac{m_{uJ}^{2}}{8\mu _{JJ}}-\frac{m_{vv}}{%
2}\right) +\frac{m_{uJ}^{2}}{8\mu _{JJ}}+\frac{m_{vv}}{2}\right] u^{2} 
\notag \\
&&+\left[ \frac{m_{uJ}^{2}}{4\mu _{JJ}}\sin ^{2}\lambda +m_{vv}\cos
^{2}\lambda \right] v^{2}+\sin (2\lambda )\left( \frac{m_{uJ}^{2}}{4\mu _{JJ}%
}-m_{vv}\right) uv.  \notag
\end{eqnarray}%
Here we are free to chose the time-dependent function $\lambda (t)$. As
previously pointed out for non-Hermitian systems with time-dependent metric,
one needs to distinguish between the Hamiltonian, that is a non-observable
operator, and the observable energy operator. This feature remains also true
when the non-Hermitian Hamiltonian is time-independent, but the metric is
dependent on time. In reverse, it simply means that when one identifies the
non-Hermitian Hamiltonian with the energy operator one has made the choice
for the metric to be time-independent. With $\eta (t)$ as specified in (\ref%
{etas}), the energy operator is computed with the help of (\ref{auxen}) to 
\begin{eqnarray}
\tilde{H}(t) &=&\eta ^{-1}(t)h(t)\eta (t)=\hat{H}+i\hbar \eta
^{-1}(t)\partial _{t}\eta (t) \\
&=&m_{JJ}J^{2}+m_{v}v+m_{vv}v^{2}+im_{uJ}uJ-\text{$\dot{\lambda}J~$}-i\frac{%
m_{uJ}}{m_{JJ}}\text{$\dot{\lambda}$}u.
\end{eqnarray}%
We note that $\tilde{H}(t)$ is also $\mathcal{PT}_{2}$-symmetric when we
include $\partial _{t}\rightarrow -\partial _{t}$ into the symmetry
transformation. In order to demonstrate that this system is quasi-exactly
solvable we specify the constants in the Hamiltonian (\ref{HH}) further to $%
m_{JJ}=4$, $m_{uJ}=2(1-\beta )\zeta $, $m_{vv}=-\beta \zeta ^{2}$, $%
m_{v}=2\zeta N$ so that it becomes 
\begin{equation}
H(N,\zeta ,\beta )=4J^{2}+i2(1-\beta )\zeta uJ-\beta \zeta ^{2}v^{2}+2\zeta
Nv,\qquad \beta ,\zeta ,N\in \mathbb{R}.  \label{newH}
\end{equation}%
This Hamiltonian can be obtained from one discussed in \cite%
{fring2016unifying} by transforming $\theta \rightarrow \theta /2$, $%
J\rightarrow 2J$ in the trigonometric representation. The constants in $%
H(N,\zeta ,\beta )$ are chosen so that it exhibits an interesting double
scaling limit $\lim_{\zeta \rightarrow 0,N\rightarrow \infty }H(N,\zeta
,\beta )=4J^{2}+2gv$ when assuming that $g:=\zeta N$. In the trigonometric
representation this limiting Hamiltonian is the Mathieu Hamiltonian.

The Hermitian Hamiltonian (\ref{hh}) simplifies in this case to%
\begin{equation}
h(t,N,\zeta ,\beta )=4J^{2}-\dot{\lambda}J+\zeta \left( 2N+\beta -1\right)
\left( \cos \lambda v-\sin \lambda u\right) +\frac{\gamma ^{2}}{4}\left(
\cos \lambda u+\sin \lambda v\right) ^{2}+\beta \zeta ^{2}C  \label{hhat}
\end{equation}%
where we denoted the Casimir operator by $C:=v^{2}+u^{2}$ and abbreviated $%
\gamma :=(1+\beta )\zeta $. In the aforementioned double scaling limit we
obtain a time-dependent Hamiltonian of the form $\lim_{\zeta \rightarrow
0,N\rightarrow \infty }h(t,N,\zeta ,\beta )=4J^{2}-\dot{\lambda}J+2g\left(
\cos \lambda v-\sin \lambda u\right) $.

\subsection{Quasi-exactly solvable Lewis-Riesenfeld invariants}

The most efficient way to solve the time-dependent Dyson equation (\ref{TDDE}%
) is to use the Lewis-Riesenfeld approach \cite{Lewis69} and compute at
first the respective time-dependent invariants $I_{h}(t)$ and $I_{H}(t)$ for
the Hamiltonian $h(t)$ and $H(t)$, see \cite%
{khantoul2017invariant,maamache2017pseudo,AndTom4}, by solving the equations%
\begin{equation}
\partial _{t}I_{H}(t)=i\hbar \left[ I_{H}(t),H(t)\right] ,\quad \text{%
and\quad }\partial _{t}I_{h}(t)=i\hbar \left[ I_{h}(t),h(t)\right] \text{.}
\label{LRin}
\end{equation}%
Unlike the corresponding Hamiltonians that have to obey (\ref{TDDE}), the
invariants are related by a similarity transformation 
\begin{equation}
I_{h}(t)=\eta (t)I_{H}(t)\eta ^{-1}(t)\text{.}  \label{simhH}
\end{equation}%
Computing the eigenstates of the invariants 
\begin{equation}
I_{h}(t)\left\vert \tilde{\phi}(t)\right\rangle =\Lambda \left\vert \tilde{%
\phi}(t)\right\rangle ,~~~~~~~I_{H}(t)\left\vert \tilde{\psi}%
(t)\right\rangle =\Lambda \left\vert \tilde{\psi}(t)\right\rangle ,~~~~~~~%
\text{with }\dot{\Lambda}=0  \label{LR1}
\end{equation}%
the solutions to the time-dependent Schr\"{o}dinger equations for $%
\left\vert \phi (t)\right\rangle $, $\left\vert \psi (t)\right\rangle $ are
simply related by a phase factor to the eigenstates of the invariants $%
~\left\vert \phi (t)\right\rangle =e^{i\alpha _{h}(t)/\hbar }\left\vert 
\tilde{\phi}(t)\right\rangle $, $\left\vert \psi (t)\right\rangle
=e^{i\alpha _{H}(t)/\hbar }\left\vert \tilde{\psi}(t)\right\rangle $. It is
easy to to derive that the two phase factors have to be identical $\alpha
_{h}=\alpha _{H}=\alpha $. They can be determined from 
\begin{equation}
\dot{\alpha}=\left\langle \tilde{\phi}(t)\right\vert i\hbar \partial
_{t}-h(t)\left\vert \tilde{\phi}(t)\right\rangle =\left\langle \tilde{\psi}%
(t)\right\vert \eta ^{\dagger }(t)\eta (t)\left[ i\hbar \partial _{t}-H(t)%
\right] \left\vert \tilde{\psi}(t)\right\rangle .
\end{equation}%
Taking now $H$ to be time-independent, we may assume $I_{H}=H+c\mathbb{I}$
with $c$ being some constant. The Lewis-Riesenfeld then just becomes a
dynamical phase factor 
\begin{equation}
\dot{\alpha}=\left\langle \tilde{\psi}\right\vert \rho (t)\left[ i\hbar
\partial _{t}-H\right] \left\vert \tilde{\psi}\right\rangle =\left\langle 
\tilde{\psi}\right\vert \rho (t)\left[ c\mathbb{I-}I_{H}\right] \left\vert 
\tilde{\psi}\right\rangle =c-\Lambda =-E,
\end{equation}%
such that $\alpha (t)=-Et$.

Next we quasi-exactly construct the Lewis-Riesenfeld invariants together
with its eigenstates for the time-dependent Hermitian and time-independent
non-Hermitian systems (\ref{hh}) and (\ref{HH}), respectively.

\subsubsection{The quasi-exactly solvable symmetry operator $I_{\hat{H}}$}

We make a general Ansatz for the invariant of $\hat{H}$ of the form%
\begin{equation}
I_{\hat{H}}=\nu _{JJ}J^{2}+\nu _{J}J+\nu _{u}u+\nu _{v}v+\nu _{uJ}uJ+\nu
_{vJ}vJ+\nu _{uu}u^{2}+\nu _{vv}v^{2}+\nu _{uv}uv,  \label{IH}
\end{equation}%
with unknown constants $\nu _{i}$. The invariant for the time-independent
system is of course just a symmetry and we only need to compute the
commutator of $I_{\hat{H}}$ with $\hat{H}$ to determine the coefficients in (%
\ref{IH}). We find the most general symmetry or invariant to be 
\begin{eqnarray}
I_{\hat{H}} &=&\nu _{JJ}J^{2}+m_{v}\frac{\nu _{JJ}}{m_{JJ}}v+im_{uJ}\frac{%
\nu _{JJ}}{m_{JJ}}uJ+\left( \nu _{vv}-m_{vv}\frac{\nu _{JJ}}{m_{JJ}}\right)
u^{2}+\nu _{vv}v^{2}  \label{IH2} \\
&=&\hat{H}+(\beta \zeta ^{2}+\nu _{vv})C,
\end{eqnarray}%
where in the last equation we have taken $\nu _{JJ}=m_{JJ}$. Since the last
term only produces an overall shift in the spectrum we set $\nu _{vv}=0$ for
convenience.

Next we compute the eigensystem for $I_{\hat{H}}$ by solving (\ref{LR1}).
Assuming the two linear independent eigenfunctions to be of the general
forms 
\begin{equation}
\tilde{\psi}_{\hat{H}}^{c}(\theta )=\psi _{0}\sum_{n=0}^{\infty
}c_{n}P_{n}(\Lambda )\cos (n\theta ),\quad \text{and\quad }\tilde{\psi}_{%
\hat{H}}^{s}(\theta )=\psi _{0}\sum_{n=1}^{\infty }c_{n}Q_{n}(\Lambda )\sin
(n\theta ),  \label{FS}
\end{equation}%
with constants $c_{n}=1/\zeta ^{n}(N+\beta )(1+\beta )^{n-1}\left[
(1+N+2\beta )/(1+\beta )\right] _{n-1}$ where $\left[ a\right] _{n}:=\Gamma
\left( a+n\right) /\Gamma \left( a\right) $ denotes the Pochhammer symbol.
The ground state $\psi _{0}=e^{-\frac{1}{2}\zeta \cos (\theta )}$ is taken
to be $\mathcal{PT}_{2}$-symmetric. The constants $c_{n}$ are chosen
conveniently to ensure the simplicity of the polynomials $P_{n}(\Lambda )$, $%
Q_{n}(\Lambda )$ in the eigenvalues $\Lambda $. We then find that the
functions $\tilde{\psi}_{\hat{H}}^{c}$ and $\tilde{\psi}_{\hat{H}}^{s}$
satisfy the eigenvalue equation provided the coefficient functions $%
P_{n}(\Lambda )$ and $Q_{n}(\Lambda )$ obey the three-term recurrence
relations%
\begin{eqnarray}
P_{2} &=&(\Lambda -4)P_{1}+\mathbf{2}\zeta ^{2}\left( N-1\right) \left(
N+\beta \right) P_{0},  \label{r1} \\
P_{n+1} &=&(\Lambda -4n^{2})P_{n}-\zeta ^{2}\left[ N+n\beta +(n-1)\right] %
\left[ N-(n-1)\beta -n\right] P_{n-1},  \label{r2} \\
Q_{2} &=&(\Lambda -4)Q_{1},  \label{r3} \\
Q_{m+1} &=&(\Lambda -4m^{2})Q_{m}-\zeta ^{2}\left[ N+m\beta +(m-1)\right] %
\left[ N-(m-1)\beta -m\right] Q_{m-1},~~~~~  \label{r4}
\end{eqnarray}%
for $n=0,2,\ldots $ and for $m=2,3,4,\ldots $ Setting $P_{0}=1$ and $Q_{1}=1$%
, the first solutions for (\ref{r1}) - (\ref{r4}) are found to be 
\begin{eqnarray}
P_{1} &=&\Lambda , \\
P_{2} &=&\Lambda ^{2}-4\Lambda -2\zeta ^{2}(N-1)(\beta +N),  \notag \\
P_{3} &=&\Lambda ^{3}-20\Lambda ^{2}+\left[ \zeta ^{2}\left( 2\beta
^{2}+7\beta -3N^{2}-3(\beta -1)N+2\right) +64\right] \Lambda +32\zeta
^{2}(N-1)(\beta +N),  \notag
\end{eqnarray}%
and%
\begin{eqnarray}
Q_{2} &=&\left( \Lambda -4\right) , \\
Q_{3} &=&(\Lambda -20)\Lambda +\zeta ^{2}(\beta -N+2)(2\beta +N+1)+64, 
\notag \\
Q_{4} &=&\Lambda ^{3}-56\Lambda ^{2}+\left[ 2\zeta ^{2}\left( 4\beta
^{2}+9\beta -N^{2}-\beta N+N+4\right) +784\right] \Lambda   \notag \\
&&+8\zeta ^{2}\left[ 5N^{2}+5(\beta -1)N-12-\beta (12\beta +29)\right] -2304.
\notag
\end{eqnarray}%
The well-known and crucial feature responsible for a system to be
quasi-exactly solvable is the occurrence of the three-term recurrence
relations and that they can be forced to terminate at certain values of $n$.
This is indeed the case and for our relations (\ref{r2}), (\ref{r4}) and can
be achieved for some specific values $n=\hat{n}$ or $m=\hat{n}$,
respectively. To see this we take $N=\hat{n}+(\hat{n}-1)\beta $ and note
that the polynomials $P_{n}$ and $Q_{m}$ factorize for $n\geq \hat{n}$, $%
m\geq \hat{n}$ as%
\begin{equation}
P_{\hat{n}+\ell }=P_{\hat{n}}R_{\ell }\qquad \text{and\qquad }Q_{\hat{n}%
+\ell }=Q_{\hat{n}}R_{\ell }\text{,}  \label{fact}
\end{equation}%
where the first $R_{\ell }$-polynomials are%
\begin{eqnarray}
R_{1} &=&\Lambda -4\hat{n}^{2}, \\
R_{2} &=&16\hat{n}^{2}(\hat{n}+1)^{2}+\Lambda \left[ \Lambda -4-8\hat{n}(%
\hat{n}+1)\right] +2\hat{n}\gamma ^{2}.
\end{eqnarray}%
Since according to (\ref{fact}) the polynomials $P_{\hat{n}}$ and $Q_{\hat{n}%
}$ are factor in all $P_{n}$ and $Q_{m}$ for $n\geq \hat{n}$ and $m\geq \hat{%
n}$, respectively, all higher order polynomial vanish when setting $P_{\hat{n%
}}(\Lambda )=Q_{\hat{n}}(\Lambda )=0$. These latter constraints are the
quantization conditions for $\Lambda $. Thus setting $P_{\hat{n}}(\Lambda )=0
$ at the different levels $\hat{n}$, we find the real eigenvalues%
\begin{eqnarray}
\hat{n} &=&1:\quad \Lambda _{1}^{c}=0, \\
\hat{n} &=&2:\quad \Lambda _{2}^{c,\pm }=2\pm 2\sqrt{1+\gamma ^{2}}, \\
\hat{n} &=&3:\quad \Lambda _{3}^{c,\ell =0,\pm 1}=\frac{4}{3}\left\{
5+2\kappa \cos \left[ \frac{\ell \pi }{3}-\frac{1}{3}\arccos \left( \frac{%
35-18\gamma ^{2}}{\kappa ^{3}}\right) \right] \right\} ,
\end{eqnarray}%
with $\kappa =\sqrt{13+3\gamma ^{2}}$, and from $Q_{\hat{n}}(\Lambda )=0$ we
find the real eigenvalues%
\begin{eqnarray}
\hat{n} &=&2:\quad \Lambda _{2}^{s}=4, \\
\hat{n} &=&3:\quad \Lambda _{3}^{s,\pm }=10\pm 2\sqrt{9+\gamma ^{2}}, \\
\hat{n} &=&4:\quad \Lambda _{4}^{s,\ell =0,\pm 1}=\frac{8}{3}\left\{ 7+%
\tilde{\kappa}\cos \left[ \frac{\ell \pi }{3}-\frac{1}{3}\arccos \left( 
\frac{143-18\gamma ^{2}}{\tilde{\kappa}^{3}}\right) \right] \right\} ,
\end{eqnarray}%
with $\tilde{\kappa}=\sqrt{49+3\gamma ^{2}}$.

Thus $\hat{H}$ is a QES system with eigenfunctions identical to those in (%
\ref{FS}) and energies $E=\Lambda -\beta \zeta ^{2}$.

\subsubsection{The quasi-exactly solvable invariant $I_{\hat{h}}$}

Next we construct the invariant $I_{\hat{h}}$ together with their
eigenfunctions. In principle we have to solve the second equation in (\ref%
{LRin}) for this purpose, however, since we already know the Dyson map we
can simply use (\ref{simhH}) and act adjointly with $\eta (t)$, as given in (%
\ref{etas}), on $I_{\hat{H}}$ as specified in (\ref{IH2}). This yields the
time-dependent invariant for $\hat{h}(t)$ as%
\begin{equation}
I_{\hat{h}}=\eta (t)I_{\hat{H}}(t)\eta ^{-1}(t)=\hat{h}+\dot{\lambda}J+\beta
\zeta ^{2}C  \label{Ih}
\end{equation}%
We convince ourselves that the relation (\ref{LRin}) is indeed satisfied by $%
I_{\hat{h}}$ as given in (\ref{Ih}) and $\hat{h}(t)$ as in (\ref{hhat}). The
eigenfunctions for $I_{\hat{h}}$ are then simply obtained as $\tilde{\phi}%
=\eta \tilde{\psi}$. From (\ref{FS}) we compute 
\begin{equation}
\tilde{\phi}_{\hat{h}}^{c}(\theta )=\phi _{0}\sum_{n=0}^{\infty
}c_{n}P_{n}(\Lambda )\cos \left[ n(\theta +\lambda )\right] ,\quad \tilde{%
\phi}_{\hat{h}}^{s}(\theta )=\phi _{0}\sum_{n=1}^{\infty }c_{n}Q_{n}(\Lambda
)\sin \left[ n(\theta +\lambda )\right] .  \label{sumh}
\end{equation}%
with ground state wavefunction $\phi _{0}=e^{-\frac{1}{4}\zeta (1+\beta
)\cos (\theta +\lambda )}$ and coefficients $c_{n}$, $P_{n}(\Lambda )$, $%
Q_{n}(\Lambda )$ as defined above. According to the above arguments, the
solutions to the time-dependent Schr\"{o}dinger equation are $\phi _{\hat{h}%
}^{c,s}(\theta )=e^{-iEt/\hbar }\tilde{\phi}_{\hat{h}}^{c,s}(\theta )$.

\subsection{A time-denpedent three level system}

For each integer value of $\hat{n}$ we have now obtained a time-dependent
QES system with a finite dimensional Hilbert space. Since it is the easiest
non-trivial example and time-dependent three-level systems are of some
interest in the literature \cite{hioe1981n,hioe1983dynamic,naudts2011} we
present here the case for $\hat{n}=2$ in more detail. From (\ref{sumh}) we
obtain three orthonormal wavefunctions%
\begin{eqnarray}
\phi _{\pm }(\theta ,t) &=&\frac{\sqrt{\gamma }}{2\sqrt{\pi N_{\pm }}}e^{-%
\frac{1}{4}\gamma \cos \left[ \theta +\lambda (t)\right] -iE_{\pm }t}\left[
\gamma +(1\pm \sqrt{1+\gamma ^{2}})\right] \cos \left[ \theta +\lambda (t)%
\right] ,  \label{f1} \\
\phi _{0}(\theta ,t) &=&\frac{\sqrt{\gamma }}{2\sqrt{\pi N_{0}}}e^{-\frac{1}{%
4}\gamma \cos \left[ \theta +\lambda (t)\right] -iE_{0}t}\sin \left[ \theta
+\lambda (t)\right] ,  \label{f2}
\end{eqnarray}%
with normalization constants%
\begin{eqnarray}
N_{\pm } &=&\gamma \left( 1+\gamma ^{2}\pm \sqrt{1+\gamma ^{2}}\right)
I_{0}\left( \gamma /2\right) -\left[ 2+2\gamma ^{2}\pm (2+\gamma ^{2})\sqrt{%
1+\gamma ^{2}}\right] I_{1}\left( \gamma /2\right) , \\
N_{0} &=&I_{1}\left( \gamma /2\right) ,
\end{eqnarray}%
and eigenenergies $E_{0}=4-\beta \zeta ^{2}$, $E_{\pm }=2-\beta \zeta
^{2}\pm 2\sqrt{1+\gamma ^{2}}$. The $I_{n}\left( z\right) $ denote here the
modified Bessel function of the first kind. The functions in (\ref{f1}) and (%
\ref{f2}) solve the time-dependent Schr\"{o}dinger equation for $\hat{h}(t)$
and are orthonormal on any interval $[\theta _{0},\theta _{0}+2\pi ]$%
\begin{equation}
\left\langle \phi _{n}(\theta ,t)\right. \left\vert \phi _{m}(\theta
,t)\right\rangle =:\dint\nolimits_{\theta _{0}}^{\theta _{0}+2\pi }\phi
_{n}^{\ast }(\theta ,t)\phi _{m}(\theta ,t)d\theta =\delta _{n,m}\qquad
n,m\in \{0,\pm \}.
\end{equation}%
We may now compute analytically all time-dependent quantities of physical
interest. For instance, the expectation values for the generators in the
trigonometric representation result to%
\begin{eqnarray}
\left\langle \phi _{\pm }(\theta ,t)\right\vert u\left\vert \phi _{\pm
}(\theta ,t)\right\rangle  &=&-\frac{M_{\pm }}{N_{\pm }}\sin \left[ \lambda
(t)\right] ,~~\left\langle \phi _{0}(\theta ,t)\right\vert u\left\vert \phi
_{0}(\theta ,t)\right\rangle =\frac{I_{2}\left( \gamma /2\right) }{%
I_{1}\left( \gamma /2\right) }\sin \left[ \lambda (t)\right] ,~~~~~~~~ \\
\left\langle \phi _{\pm }(\theta ,t)\right\vert v\left\vert \phi _{\pm
}(\theta ,t)\right\rangle  &=&\frac{M_{\pm }}{N_{\pm }}\cos \left[ \lambda
(t)\right] ,~~\left\langle \phi _{0}(\theta ,t)\right\vert v\left\vert \phi
_{0}(\theta ,t)\right\rangle =-\frac{I_{2}\left( \gamma /2\right) }{%
I_{1}\left( \gamma /2\right) }\cos \left[ \lambda (t)\right] , \\
\left\langle \phi _{\ell }(\theta ,t)\right\vert J\left\vert \phi _{\ell
}(\theta ,t)\right\rangle  &=&0,~~~~\ell \in \{0,\pm \},
\end{eqnarray}%
where we abbreviated%
\begin{equation}
M_{\pm }=\gamma \left( 1-\gamma ^{2}\pm \sqrt{1+\gamma ^{2}}\right)
I_{1}\left( \gamma /2\right) +\left[ 2+2\gamma ^{2}\pm (2+\gamma ^{2})\sqrt{%
1+\gamma ^{2}}\right] I_{2}\left( \gamma /2\right) .
\end{equation}%
Similarly we may obtain any kind of $n$-level system from  (\ref{sumh}).

\section{Conclusions}

We have provided new analytical solutions for the time-dependent Dyson
equation. The time-dependent non-Hermitian Hamiltonians (\ref{H}) considered
are expressed in terms linear and bilinear combinations of the generators
for an Euclidean $E_{2}$-algebra respecting the $\mathcal{PT}_{i}$%
-symmetries defined in (\ref{coe}). Restricting the coefficient functions
appropriately, the corresponding time-dependent Hermitian Hamiltonians were
constructed. We expect a different qualitative behaviour for Hamiltonians
belonging to different symmetry classes.

A specific $\mathcal{PT}_{2}$-symmetric system was analyzed in more detail.
For that model we assumed the non-Hermitian Hamiltonian to be
time-independent so that we could employ the metric picture. This enabled us
to compute the corresponding eigensystems in a quasi-exactly solvable
fashion using Lewis-Riesenfeld invariants. Thus we found for the first time
quasi-exactly solvable systems for Hamiltonians with explicit
time-dependence. A time-dependent Hermitian three-level system is presented
in more detail.

Evidently there are many open issues and problems for further investigations
left. Having solved the time-dependent Dyson equation for a large class of
models in section 2, it would be interesting to solve their corresponding
time-dependent Schr\"{o}dinger equation as carried out for the model in
section 3. Furthermore, it is desirable in this type of analysis to allow an
explicit time-dependence also in the non-Hermitian Hamiltonians. Clearly one
may also generalize these studies to Euclidean algebras of higher rank and
other types of Lie algebras.\medskip 

\noindent \textbf{Acknowledgments:} TF is supported by a City, University of
London Research Fellowship.

\newif\ifabfull\abfulltrue


\begin{thebibliography}{10}

\bibitem{OP2}
M.~A. Olshanetsky and A.~M. Perelomov,
\newblock Classical integrable finite dimensional systems related to Lie
  algebras,
\newblock Phys. Rept. {\bf 71}, 313--400 (1981).

\bibitem{OP4}
M.~A. Olshanetsky and A.~M. Perelomov,
\newblock Quantum integrable systems related to Lie algebras,
\newblock Phys. Rept. {\bf 94}, 313--404 (1983).

\bibitem{Turbiner00}
A.~V. Turbiner,
\newblock Quasi-Exactly-Solvable problems and sl(2) Algebra,
\newblock Commun. Math. Phys. {\bf 118}, 467--474 (1988).

\bibitem{Tur0}
A.~Turbiner,
\newblock Lie algebras and linear operators with invariant subspaces,
\newblock Lie Algebras, Cohomologies and New Findings in Quantum Mechanics,
  Contemp. Math. AMS, (eds N. Kamran and P.J. Olver) {\bf 160}, 263--310
  (1994).

\bibitem{Hum}
J.~E. Humphreys,
\newblock Introduction to Lie Algebras and Representation Theory,
\newblock Springer, Berlin  (1972).

\bibitem{E2Fring}
A.~Fring,
\newblock E2-quasi-exact solvability for non-Hermitian models,
\newblock J. Phys. {\bf A48}, 145301(19) (2015).

\bibitem{E2Fring2}
A.~Fring,
\newblock A new non-Hermitian E2-quasi-exactly solvable model,
\newblock Phys. Lett. {\bf 379}, 873--876 (2015).

\bibitem{fring2016unifying}
A.~Fring,
\newblock A unifying E2-quasi exactly solvable model,
\newblock in {\em In: Bagarello F., Passante R., Trapani C. (eds) Non-Hermitian
  Hamiltonians in Quantum Physics. Springer Proceedings in Physics, vol 184.
  Springer, Cham}, pages 235--248, Springer, 2016.

\bibitem{mayer2000time}
D.~Mayer, A.~Ushveridze, and Z.~Walczak,
\newblock On time-dependent quasi-exactly solvable problems,
\newblock Mod. Phys. Lett. A {\bf 15}(19), 1243--1251 (2000).

\bibitem{hou1999quasi}
X.~Hou and M.~Shifman,
\newblock A quasi-exactly solvable N-body problem with the sl (N+ 1) algebraic
  structure,
\newblock Int. J. Mod. Phys. A {\bf 14}(19), 2993--3003 (1999).

\bibitem{AndTom1}
A.~Fring and T.~Frith,
\newblock Exact analytical solutions for time-dependent Hermitian Hamiltonian
  systems from static unobservable non-Hermitian Hamiltonians,
\newblock Phys. Rev. A {\bf 95}, 010102(R) (2017).

\bibitem{AndTom2}
A.~Fring and T.~Frith,
\newblock Metric versus observable operator representation, higher spin models,
\newblock Eur. Phys. J. Plus , 133: 57 (2018).

\bibitem{Urubu}
F.~G. Scholtz, H.~B. Geyer, and F.~Hahne,
\newblock Quasi-Hermitian Operators in Quantum Mechanics and the Variational
  Principle,
\newblock Ann. Phys. {\bf 213}, 74--101 (1992).

\bibitem{Benderrev}
C.~M. Bender,
\newblock Making sense of non-Hermitian Hamiltonians,
\newblock Rept. Prog. Phys. {\bf 70}, 947--1018 (2007).

\bibitem{Alirev}
A.~Mostafazadeh,
\newblock Pseudo-Hermitian Representation of Quantum Mechanics,
\newblock Int. J. Geom. Meth. Mod. Phys. {\bf 7}, 1191--1306 (2010).

\bibitem{CA}
C.~Figueira~de Morisson~Faria and A.~Fring,
\newblock Time evolution of non-Hermitian Hamiltonian systems,
\newblock J. Phys. {\bf A39}, 9269--9289 (2006).

\bibitem{time1}
A.~Mostafazadeh,
\newblock Time-dependent pseudo-Hermitian Hamiltonians defining a unitary
  quantum system and uniqueness of the metric operator,
\newblock Physics Letters B {\bf 650}(2), 208--212 (2007).

\bibitem{time6}
M.~Znojil,
\newblock Time-dependent version of crypto-Hermitian quantum theory,
\newblock Physical Review D {\bf 78}(8), 085003 (2008).

\bibitem{time7}
J.~Gong and Q.-H. Wang,
\newblock Time-dependent PT-symmetric quantum mechanics,
\newblock J. Phys. A: Math. and Theor. {\bf 46}(48), 485302 (2013).

\bibitem{fringmoussa}
A.~Fring and M.~H.~Y. Moussa,
\newblock Unitary quantum evolution for time-dependent quasi-Hermitian systems
  with nonobservable Hamiltonians,
\newblock Physical Review A {\bf 93}(4), 042114 (2016).

\bibitem{AndTom3}
A.~Fring and T.~Frith,
\newblock Mending the broken PT-regime via an explicit time-dependent Dyson
  map,
\newblock Phys. Lett. A , 2318 (2017).

\bibitem{AndTom4}
A.~Fring and T.~Frith,
\newblock Solvable two-dimensional time-dependent non-Hermitian quantum systems
  with infinite dimensional Hilbert space in the broken PT-regime,
\newblock J. of Phys. A: Math. and Theor. {\bf 51}(26), 265301 (2018).

\bibitem{mostafazadeh2018energy}
A.~Mostafazadeh,
\newblock Energy Observable for a Quantum System with a Dynamical Hilbert Space
  and a Global Geometric Extension of Quantum Theory,
\newblock arXiv preprint arXiv:1803.04175  (2018).

\bibitem{EW}
E.~Wigner,
\newblock Normal form of antiunitary operators,
\newblock J. Math. Phys. {\bf 1}, 409--413 (1960).

\bibitem{DFM}
S.~Dey, A.~Fring, and T.~Mathanaranjan,
\newblock Non-Hermitian systems of Euclidean Lie algebraic type with real
  eigenvalue spectra,
\newblock Annals of Physics {\bf 346}, 28--41 (2014).

\bibitem{DFM2}
S.~Dey, A.~Fring, and T.~Mathanaranjan,
\newblock Spontaneous PT-symmetry breaking for systems of noncommutative
  Euclidean Lie algebraic type,
\newblock Int. J. of Theor. Phys. {\bf 54}(11), 4027--4033 (2015).

\bibitem{Lewis69}
H.~Lewis and W.~Riesenfeld,
\newblock An Exact quantum theory of the time dependent harmonic oscillator and
  of a charged particle time dependent electromagnetic field,
\newblock J. Math. Phys. {\bf 10}, 1458--1473 (1969).

\bibitem{khantoul2017invariant}
B.~Khantoul, A.~Bounames, and M.~Maamache,
\newblock On the invariant method for the time-dependent non-Hermitian
  Hamiltonians,
\newblock The European Physical Journal Plus {\bf 132}(6), 258 (2017).

\bibitem{maamache2017pseudo}
M.~Maamache, O.~K. Djeghiour, N.~Mana, and W.~Koussa,
\newblock Pseudo-invariants theory and real phases for systems with
  non-Hermitian time-dependent Hamiltonians,
\newblock The European Physical Journal Plus {\bf 132}(9), 383 (2017).

\bibitem{hioe1981n}
F.~T. Hioe and J.~H. Eberly,
\newblock N-level coherence vector and higher conservation laws in quantum
  optics and quantum mechanics,
\newblock Phys. Rev. Lett. {\bf 47}(12), 838 (1981).

\bibitem{hioe1983dynamic}
F.~Hioe,
\newblock Dynamic symmetries in quantum electronics,
\newblock Phys. Rev. A {\bf 28}(2), 879 (1983).

\bibitem{naudts2011}
J.~Naudts and W.~O. de~Galway,
\newblock Analytic solutions for a three-level system in a time-dependent
  field,
\newblock Physica D: Nonlinear Phenomena {\bf 240}(6), 542--545 (2011).

\end{thebibliography}
\end{document}